\documentclass[10pt,a4paper,twocolumn]{vupreprint}
\usepackage[bindingoffset=0.5cm,textheight=25cm,hdivide={2.5cm,*,3cm}, vdivide={*,24cm,*}]{geometry}
\usepackage{amsmath}
\usepackage{amsfonts}
\usepackage{amssymb}
 \usepackage[singlelinecheck=true,font=small]{caption}
 \usepackage[numbers,square,comma,sort&compress]{natbib}
 \usepackage{fancyhdr}
 \usepackage{fancybox}
 \usepackage{authblk}
 \usepackage{ftnright}
\usepackage[pdftex]{graphicx}
\usepackage[pdftex,bookmarksnumbered=true,breaklinks=true]{hyperref}
\usepackage{subfig}
\usepackage{booktabs}
\newcommand{\eqnref}[1]{Eqn.~(\ref{#1})}		
\newcommand{\figref}[1]{Fig.~\ref{#1}}			
\newcommand{\tabref}[1]{Tab.~\ref{#1}}			
\newcommand{\secref}[1]{Section~\ref{#1}}		
\renewcommand{\a}{\alpha}

\renewcommand{\l}{\lambda}
\newcommand{\m}{\mu}

\newcommand{\s}{\sigma}

\newcommand{\G}{\Gamma}

\renewcommand{\Xi}{\Xi}

\newcommand{\inv}[1]{\frac{1}{#1}}					
 \doi{10.1103/PhysRevB.88.165429}
 \journalref{R. I. P. Sedmik, A. Almasi, and D. Iannuzzi}{Phys. Rev. B}{88}{165429}{2013}{The locality of surface interactions on colloidal probes}
 \corrauthtext{{Corresponding author email: \url{r.sedmik@vu.nl}}}
  \author{R. I. P. Sedmik\footnote{xxx}}
  \author{A. Almasi}
  \author{D. Iannuzzi}
 \affil{Department of Physics \& Astronomy and LaserLaB, VU Amsterdam (Netherlands)}
\hypersetup{pdfauthor={\theauthors},pdftitle={\thetitle}}
\graphicspath{{./}{./figures/}}
 
\begin{document}
\twocolumn[\begin{@twocolumnfalse}%

\maketitle
\footnotetext{Corresponding author: r.sedmik@vu.nl}
%
%
\begin{abstract}
The Casimir and electromagnetic interactions between objects at short separations are strongly influenced by the local geometry near the point of closest approach. In this paper we demonstrate that the assumptions underlying common statistical analysis of roughness may not hold in experiments using micro-spheres as interacting objects. Based on an extensive experimental and numerical analysis of the surface topology of the widely used colloidal 
particle types 4310A and 4320A, we show that the actual variation in the local surface curvature may give rise to large uncertainties in the comparison of experimental data to theories.
\end{abstract}
\end{@twocolumnfalse}]
\thisfancyput(-12pt,-719pt){\parbox{8cm}{\flushleft${}^\ast$\small\thecorrauthtext}}
%
%
\section{Introduction}
Over the past two decades hydrodynamic~\cite{Vinogradova:2011} and Casimir~\cite{Bordag:2009a,Klimchitskaya:2009cw,Rodriguez:2011} (or van der Waals) interactions at short surface separations have received increased attention. Numerous measurements using a wide variety of experimental setups have been conducted in order to better understand these effects. In particular, the technical maturity and availability of atomic force microscopes (AFM) has made these instruments popular for the measurement of forces at short surface separations. As both the hydrodynamic and the Casimir forces scale with distance as well as with the area of the opposing surfaces, one would ideally use a parallel plate configuration in order to maximize the force. Due to the technical difficulty of maintaining parallelism, however, measurements are mostly performed in geometries with higher degrees of symmetry such as a sphere opposing a flat plate or orthogonal cylinders.\\
When comparing experimental results to theoretical predictions, the imperfections of the experimental surfaces have to be taken into account. For hydrodynamics, a wide spectrum of influences from tribological parameters, such as wettability~\cite{Cho:2004}, surface contamination~\cite{Shields:1983,Schrader:1984}, trapping of nano bubbles~\cite{Lauga:2004,Neto:2005}, and roughness~\cite{Zhu:2001,Zhu:2002,Bonaccurso:2002,Kunert:2010} have been investigated (for a recent review see \cite{Vinogradova:2011}). In the field of Casimir physics, dielectric properties~\cite{deMan:2010,Torricelli:2010,Chen:2006,Chen:2007,Banishev:2012}, the thickness of surface layers~\cite{lisanti:2005}, local variations in the Fermi potential (patches)~\cite{Speake:2003,Kim:2010,Behunin:2012}, and roughness~\cite{Klimchitskaya:1999,MaiaNeto:2005,vanZwol:2008b,Broer:2012} have been shown to influence the measured forces significantly. Stochastic irregularities also seem to play an important role for capillary forces~\cite{vanZwol:2007b} and the mechanical characteristics of micro-electromechanical switches~\cite{Palasantzas:2005}.\\
Being short ranged, hydrodynamic slip effects as well as the Casimir force crucially depend on the geometrical properties of the interacting objects. In most studies, geometry is accounted for by considering the global radius of curvature and the roughness amplitude~\cite{Neto:2001}. While these two parameters represent the extreme scales of surface geometries, the intermediate scale, namely local variations in the curvature (sometimes referred to as `waviness'), are mostly neglected in studies utilizing spherical (or cylindrical) probes. However, theoretical studies~\cite{Bezerra:2011} have shown that precision Casimir experiments performed with lenses of centimeter-size radius are highly sensitive for non-sphericity. Other authors report variations of several tens of percents in the hydrodynamic drainage force between a sphere and a plate for local dents~\cite{Zhu:2011a}, large-scale asperities~\cite{Fan:2005}, and generic stochastic variations of the surface topology~\cite{Guriyanova:2010}.\\
It is the aim of the present study to demonstrate that the basic assumptions of commonly applied statistical methods for the description of roughness (and general corrugations) of spheres may not always be met. Based on a series of atomic force microscope (AFM) measurements on commonly used colloidal probes, we derive global and local values for the surface curvature. At hand of experimental topology data we show that the Casimir force may vary by several percent in dependence on the position of the point of closest approach on the sphere -- an effect which cannot be covered statistically.\\
The paper is organized as follows.
\secref{sec:locality} gives a theoretical and numerical assessment of the effective area on colloid surfaces that give the major contribution to the Casimir and electrostatic interactions in an experiment. 
In \secref{sec:methods} we describe the experimental methods used in the present work. A detailed presentation and analysis of the data on radii and variations of the Casimir force in \secref{sec:results} is followed by a brief summary and conclusion of our work in \secref{sec:concl}.
\section{Locality of surface interactions}
\label{sec:locality}
\begin{figure}[ht]
\centering
   \includegraphics{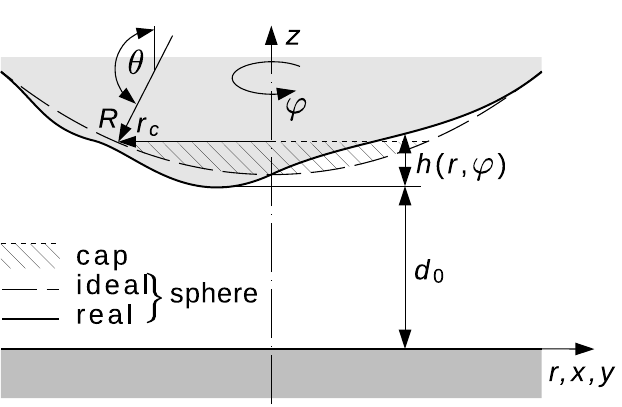}
\caption{Geometry of a corrugated sphere opposing an ideally flat plate (not to scale).\label{fig:geom}}
\end{figure}
\begin{figure}[!ht]
\centering
   \includegraphics{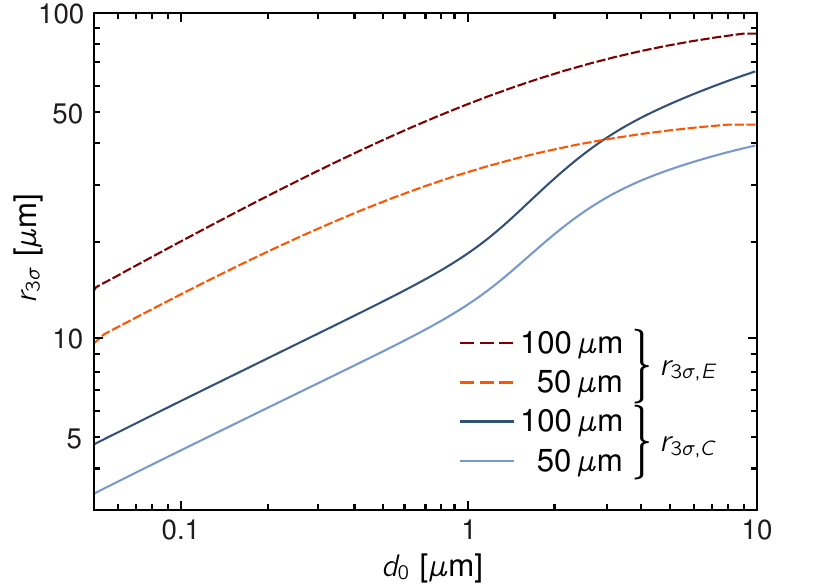}
\caption{(color online) Dependence of the effective cap radii $r_{3\sigma,C}$ and $r_{3\sigma,E}$ for Casimir and electrostatic interactions on distance $d$, for particles of radii $R=50\,\m{\rm m}$ and $R=100\,\m$m.\label{fig:r95_ddep}}
\end{figure}

Under the sole premise that the smallest curvature $R$ of the interacting surfaces is much larger than the separation $d_0$ at the point of closest approach ($d_0\ll R$), the Casimir force can be estimated by the Derjaguin approximation (also known as proximity force approximation (PFA))~\cite{Derjaguin:1956}. In order to quantify the `locality' of the interaction it is instructive to compute the size of the cap on an ideal sphere that contributes most to the force between the same sphere and a flat plate. For this purpose, we write the PFA in the form
\begin{align}
 \label{eq:pfa}
 F_C(r_c, d)= \int\limits_0^{r_c}\!{\rm d}r\, \int\limits_0^{2\pi}{\rm d\varphi}\,r P_{C,pp}\left(d(r,\varphi)\right)\,,
\end{align}
where $d(r,\varphi)$ is the local surface separation, $r_c$ is the lateral radius of the cap (see \figref{fig:geom}), $\varphi$ is the azimuthal angle, and $P_{C,pp}(d)$ stands for the Casimir pressure for parallel plates, as computed via Lifshitz theory~\cite{lifshitz:1956}. For a perfect sphere of radius $R$, the distance function is $d(r,\varphi)=d(r)=d_0+R-\sqrt{R^2-r^2}$, and we can compute numerically the ratio $F_C(r_c,d_0)/F_C(R,d_0)$. 
A parametric analysis evaluating \eqnref{eq:pfa} for room temperature, and gold as material for all surfaces (where the dielelectric function has been computed on the basis of tabulary data and the Drude model as described for example in~\cite{Decca:2005}), $R=50\,\mu$m, and $d_0=100\,$nm, a fraction of $95.5\,$\% of the total force is generated within a radius of $r_c\equiv r_{3\sigma,C}=4.65\,\mu$m around the point of closest approach. For a sphere with $R=100\,\mu$m, a similar estimation results in $r_{3\sigma,C}=6.47\,\mu$m. \figref{fig:r95_ddep} shows the dependence of $r_{3\sigma,C}$ on distance. Although being based on an approximation, these results for $r_{3\sigma,C}$ clearly demonstrate that, in typical AFM Casimir experiments using metal-coated colloid particles, the major contribution to the measurement comes from a very small fraction of the total surface area.
We will discuss the implications of these findings with respect to roughness corrections further in \secref{sec:results_locality}.\\
The surface separation is determined in many experiments from a calibration based on electrostatic forces. Hence, it is interesting to analyze the `locality' of the latter interaction in a similar way as described above for the Casimir force. For this purpose, one can use a modified version of the widely used theory by Smythe~\cite{Smythe:1950} for the electrostatic attraction between a plate and a sphere. According to the method of mirror charges, the potential $\phi(\mathbf{x})$ in three-dimensional Cartesian coordinates $\mathbf{x}=(x,y,z)$ can be expressed as,
\begin{align}
\label{eq:potential_smythe}
 \phi(\mathbf{x})&=\inv{4\pi \varepsilon_0}\sum\limits_{j=1}^\infty\left[\frac{q_j}{\left\|\mathbf{s}_j-\mathbf{x}\right\|}+\frac{p_j}{\left\|\mathbf{r}_j-\mathbf{x}\right\|}\right]\,,\\
\text{where }\mathbf{s}_j&=\left(0,0, d_0+R\left[1-\frac{\sinh (j-1)\a}{j\sinh j\a}\right]\right)\,,\nonumber\\
q_j&=-p_j=4\pi\varepsilon_0 d_0 \left(\sqrt{\tfrac{d}{d+2R}}\sinh j\a\right)^{-1}\,,\nonumber\\
\mathbf{r}_j&=-\mathbf{s}_{j}\,,\quad \text{and }\a=\cosh^{-1}\!\left(1+\tfrac{d}{R}\right)\,.\nonumber
\end{align}
The force generated by a local cap of radius $r_c$, which is centered around the point of closest approach (see \figref{fig:geom}) is obtained from Coulomb's law by integrating the square of the electric field  $\mathbf{E}=-\nabla\phi(\mathbf{x},d_0)$ (standing parallel to the surface normal vector $\mathbf{n}$ of the sphere, with $\|\mathbf{n}\|=1$), over the cap area,
\begin{align}
\label{eq:force_es}
   F_{E}(r_c)=2\pi \bigg|\mathbf{e}_z\int\limits_\pi^{\theta(r_c)}{\rm d}\theta\, (\mathbf{n}\cdot \mathbf{E})^2R^2\sin\theta\bigg|\,,
\end{align}
where we have used spherical coordinates $(R,\theta,\varphi)$ with the relations $x=R\cos\varphi\sin\theta$, $y=R\sin\varphi\sin\theta$, $z=d_0+R(1-\cos\theta)$, the unit vector in $z$-direction $\mathbf{e}_z$ with $\|\mathbf{e}_z\|=1$, and the definition $\theta(r_c)=\pi-\sin^{-1}({r_c}/{R})$. \eqnref{eq:force_es} is amenable for numerical evaluation and allows to determine the effective radius $r_{3\sigma,E}$ for the electrostatic force from the ratio $F_E(r_c,d_0)/F_E(R,d_0)$\footnote{Note that the error from neglecting the upper half of the sphere is $\lesssim5\,$\% for $d_0<3\,\m$m, but is irrelevant at all distances due to the normalization by $F_E(R,d_0)$. The same computation with $F_{E,compl}(R, d_0)$ (obtained by setting $\theta(r_c)\to 0$ in \eqnref{eq:force_es}) instead of $F_E(R,d_0)$ yields slightly smaller values of $r_{3\sigma,E}$ at large distance.}.
The results of this analysis are collected in \figref{fig:r95_ddep}. At very large $d_0$, the effective radii contributing to both Casimir and electrostatic interactions converge to $R$. For $d_0\lesssim 1\,\m$m, $r_{3\sigma,E}$ is approximately a factor of 3 larger than $r_{3\sigma,C}$.
We note that the results obtained here for $r_{3\sigma,C}$ state a non-trivial extension of the estimates of the effective length scale given in Ref.~\cite{vanZwol:2009}, which are limited to a power-law description of the forces. 

\section{Experimental methods}
\label{sec:methods}
We have investigated the commercially available colloidal particles 4310A and 4320A from Duke Scientific, which have found application in numerous experiments using AFM. These particles have NIST-certified diameters of $201\pm3.2\,\m$m~\footnote{Note that NIST-certified values are based on only two measurements and may, thus, not be statistically representative.} (standard deviation $7.8\,\m$m) and $100\pm1.5\,(1.6)\,\m$m, respectively.\\
For both types of spheres large dense arrays of particles were fixed with epoxy glue on a glass slide. Small conglomerates of $4\times 4$ colloids with optically clean surfaces were then selected for further investigation. The surface topology was measured on several spots per sphere by using a Veeco Nanoscope III AFM in closed loop tapping mode. Then, the arrays were RF sputtered with $5\,$nm Cr followed by $100\,$nm Au, following the same procedures and parameters used for the preparation of probes for recent Casimir and hydrodynamic experiments~\cite{deMan:2009a,deMan:2009,deMan:2010,Sedmik:2013a}. All spheres were imaged in a FEI Phenom SEM, followed by a second series of measurements in the AFM. \\
For the purpose of this study, a proper calibration of the  instruments is of vital importance. Prior to the first AFM measurements we performed the full calibration procedure following the steps provided by the manufacturer using a commercially available NT-MDT TGG1 grating, which consists of a flat surface with well defined periodical triangular ridges. As a test of the calibration, and in order to give a proof for the feasibility to accurately measure the surface curvature radius, we have performed tapping mode imaging on stripped Corning SMF-28 fibers featuring a very precisely defined nominal diameter of $125\pm0.7\,\m$m with a maximum non-circularity of $0.5\,$\%. By fitting the resulting topology data of the cylindrical surfaces we obtained a value for the radius equal to $62.61\pm0.01\,(0.39)\,\m$m. This result demonstrates the fitness of the applied method to accurately determine the curvature radius of a surface. For clarity, we have to note 
that radius values obtained from surface height data throughout this work are based on unweighted least square fits (see \secref{sec:results_rad_global}). We perform statistical averaging of radii measured on the same sphere (or fiber), using the inverse square of the $68.3\,$\% confidence intervals of the parameter $R$ of the fits as weights, and report the weighted mean, its uncertainty, and the weighted standard deviation (in brackets).\\
Tapping mode scans\footnote{We would like to note that the investigation of spherical particles by means of standard tapping mode measurements has proven to be prone to approach errors, which eventually lead to damage on the probe surface. For future experiments, it is thus recommended to use the reverse imaging technique described in the literature~\cite{Neto:2001} to characterize the surface of colloidal tip AFM probes.} are performed along a rectangular raster, resulting in a fast (along the single lines of the raster, $\mathbf{x}_f$) and a slow (orthogonal to 
the lines, $\mathbf{x}_s\bot \mathbf{x}_f$) scan direction. In the evaluation of the topology of the SMF-28 samples only data have been taken into account which were acquired with the fast scanning direction oriented orthogonally to the fiber. Scanning parallel to the cylinder axis (with $\mathbf{x}_s$ being orthogonal to the fiber) resulted in deviations from the stated results due 
to the presence of drift. We found that this drift was rather constant over all our measurements and samples, yielding a distance scaling factor $f_C=0.70\pm0.06$ between calibrations performed along $\mathbf{x}_s$ and $\mathbf{x}_f$, respectively. 
A rotation of the scan direction by 90 degrees without moving the sample yielded the same value for $f$. Finally, using the TGG1 grating, we could completely exclude the appearance of drift along $\mathbf{x}_f$. Therefore, we speculate that this systematic effect is caused by an error in the scanner of our AFM. However, due to the strong reproducibility, consistency, and linearity, the drift can be removed from the data by the application of a scaling factor along $\mathbf{x}_s$ (see \secref{sec:results}).
 \begin{table*}
  \centering
 \caption{Results for the sphere radius $R$, and the roughness $\s$ (weighted averaged mean over results from scans on the same sphere type, format: RMS\{PV\}), as obtained by SEM and AFM measurements.\label{tab:global_res}}
 \setlength{\tabcolsep}{3pt}
 \begin{tabular}{c r@{.}l@{$\pm$}l r@{.}l r@{.}l r r@{.}l r@{.}l r}
  \hline\hline
type & \multicolumn{3}{c}{SEM} & \multicolumn{5}{c}{AFM before dep.} & \multicolumn{5}{c}{AFM after dep.}\\
   & \multicolumn{3}{c}{$R$ [$\m$m]} & \multicolumn{2}{c}{$R$ [$\m$m]} & \multicolumn{3}{c}{$\s$ [nm]} & \multicolumn{2}{c}{$R$ [$\m$m]} & \multicolumn{3}{c}{$\s$ [nm]}\\
   \hline
4310A&  49&8&$0.2\,(0.7)$ &  48&$93\pm0.46\,(1.3)$ & 1&7& \{35\} &47&$59\pm 0.50\,(1.4)$ & 5&7&\{137\}\\
4320A& 102&3&$0.5\,(3.6)$ & 100&$9\phantom{0}\pm2.8\,(6.9)$& 10&2&\{131\} &99&$3\phantom{0}\pm3.0\phantom{0}\,(7.3)$ & 13&2&\{167\}\\
\hline
\hline
 \end{tabular}
\end{table*}
%
%
%
\section{Analysis and Results}
\label{sec:results}
Several series of analysis have been performed on all data. First, in \secref{sec:results_rad_global} the accuracy and spread in the determination of sphere radii from AFM scans are investigated. The results are then compared to the outcomes of a more commonly used method using a SEM. In a second analysis in \secref{sec:results_locality}, we focus on the influence of locality on the Casimir force, and the local variation in the surface curvature.
\subsection{Radius determination via AFM}
\label{sec:results_rad_global}
In order to characterize the investigated sample of particles statistically, and to further verify our method by comparison of the results to the literature~\cite{Neto:2001}, the curvature radius of 32 spheres has been measured. For this purpose, AFM topology data $h_{AFM}(x,y)$ with lateral dimensions $15\times15\,\m{\rm m}^2$ have been analyzed without the application of filters or any automatic corrections in order to avoid the introduction of systematic errors. Using a standard least squares method, $h_{AFM}(x,y)$ was fit by a function
\begin{align}
 \label{eq:sph_height_function}
 h_{f}(x,y)\hspace{-1pt}=\hspace{-1pt}h_0\hspace{-1pt}+\hspace{-1pt}R\hspace{-1pt}-\hspace{-1pt}\sqrt{R^2\hspace{-1.5pt}-\hspace{-1pt}(x\hspace{-1pt}-\hspace{-1pt}x_0)^2\hspace{-1.5pt}-\hspace{-1pt}(f_D y\hspace{-1pt}-\hspace{-1pt}y_0)^2},
\end{align}
where $x$ and $y$ are the fixed lateral coordinates of the scan, $R$, $x_0$, $y_0$, and $h_0$ are free parameters representing the global radius of the sphere, and offsets in all three Cartesian coordinates, respectively. $f_D$ is a scaling parameter, which accounts for the constant drift in the slow scanning direction (see discussion in \secref{sec:methods}). Note that, because $f_D$ also is a free fit parameter, the possibly erroneous curvature along the slow scanning direction $\mathbf{y}\equiv\mathbf{x}_s$ does not influence the final result for $R$. By averaging over the outcomes of all performed data fits, we obtain a value $f_D=0.75\pm0.05$, which compares very well to the drift factor $f_C$, determined in the calibration, as mentioned in \secref{sec:methods}.\\
The root mean square (RMS) and peak-valley (PV) roughness have been measured from the fit residuals $h_{AFM}-h_f$.
In order to obtain an independent set of measurements to which the results of our AFM analysis can be compared to, we have
determined the radius of all colloids in the present study by means of circular fits to their circumference in SEM images.
 Results for the roughness and the global radius, as extracted from AFM data taken before and after the sputter coating, and from SEM data (only after the deposition) are given in \tabref{tab:global_res}.
The radii extracted from SEM measurements generally comply with the NIST certification of the spheres. In direct comparison, the results for $R$ from AFM fits before the sputter process show good agreement with the corresponding values from SEM data: $R_{\text{AFM}}/R_{\text{SEM}}-1=1.9\pm 0.3\,(2.1)\,$\% on 4310A and $0.4\pm0.4\,(5.8)\,$\% on 4320A. After the deposition of gold, a slight change (of $-5.5\pm3.0\,(2.2)\,$\%) can be observed in $R_{\text{AFM}}$ for of spheres of type 4310A, while the results for 4310A remain constant within the error of determination (relative change $0.6\pm0.5\,(10.5)\,$\%). The average roughness obtained from the same data increases after sputtering. This change is most pronounced for the (initially) relatively smooth 4310A colloids, for which the RMS value rises by a factor $\sim 3.4$, and the peak value by a factor $4.1$. With applied coating, the PV roughness of 4310A and 4320A colloid species are comparable, while the RMS value still stays clearly lower on 4310A. These observations compare well with the findings of other investigations on the same sphere types~\cite{vanZwol:2008,vanZwol:2008b}.\\
We would like to note that the method applied to extract the curvature from AFM topology data in the present work is different from the one of Refs.~\cite{Neto:2001,Zhu:2011a}. These authors used only the height profile along one single cross section of the data to derive the global radius while our fit to $h_{f}(x,y)$ is applied to the entire data set.\\
\subsection{Locality, statistics, and the Casimir force}
\label{sec:results_locality}
As discussed in \secref{sec:locality}, due to the finite curvature the actual area contributing most to the Casimir and electrostatic forces between a sphere and a plate is very small. In consequence, it is clear that the interaction measured in an experiment crucially depends on the \emph{local} topology of the surface. 
This issue has been investigated theoretically and experimentally for hydrodynamic forces in the case of a deformation (flattening) of the colloid~\cite{Zhu:2011a} as well as for asperities~\cite{Fan:2005} and patterned surfaces~\cite{Guriyanova:2010}. All studies have found variations of several tens of percent in the force. For the Casimir force, dedicated investigations~\cite{Chen:2002,Chen:2002a,Blagov:2004,Chiu:2010} have shown a significant dependence of the interaction on the relative lateral position of regular corrugations on a sphere with respect to an opposed flat surface with structures of the same periodicity. Also, it has been recognized early~\cite{Klimchitskaya:1996b} that non-stochastic large-scale deviations from perfect sphericity have a significant influence on the Casimir force.
However, for standard sphere-plate experiments reported in the literature, in the analysis of measurements the imperfections of the surfaces of spherical AFM probes are considered statistically only, either via perturbative approaches (see for example~\cite{Genet:2003,MaiaNeto:2005,vanZwol:2008}), application of the PFA~\cite{Bordag:2001,Chen:2004,Decca:2005}, or combinations thereof~\cite{Broer:2011,Broer:2012}. 
The underlying assumptions defining the applicability of statistical methods are~\cite{Genet:2003} that:
\begin{enumerate}
 \item the area associated with the lateral correlation length $\xi$ (see below and Ref.~\cite{Palasantzas:1993}) of the roughness is much smaller than the effective interaction area: $R d_0 \pi\gg \xi^2\pi$, 
 \item the roughness profile is invariant with respect to the lateral position of the sphere.
\end{enumerate}
\begin{figure*}[!ht]
\centering
   \subfloat[Height auto-correlation function]{\raisebox{-0.1pt}{\includegraphics{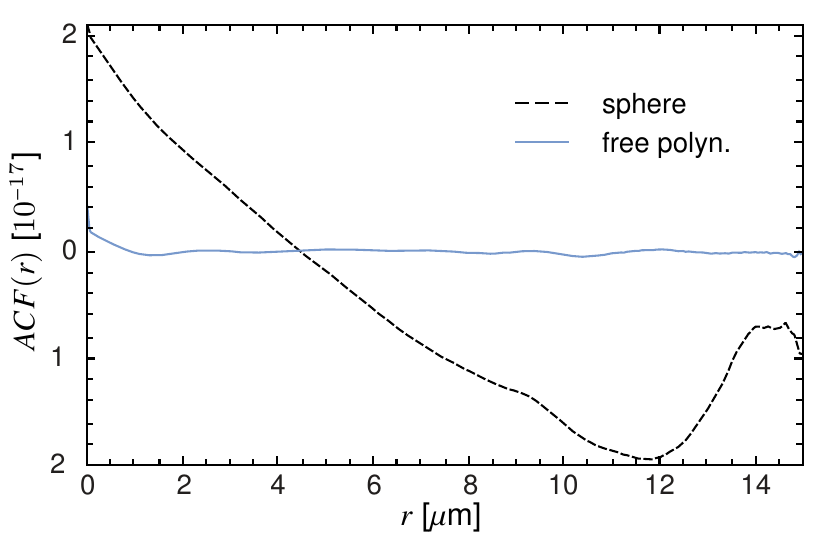}\label{fig:height_stats_ac_fun}}}
   \subfloat[Histogram and cumulative height]{\includegraphics{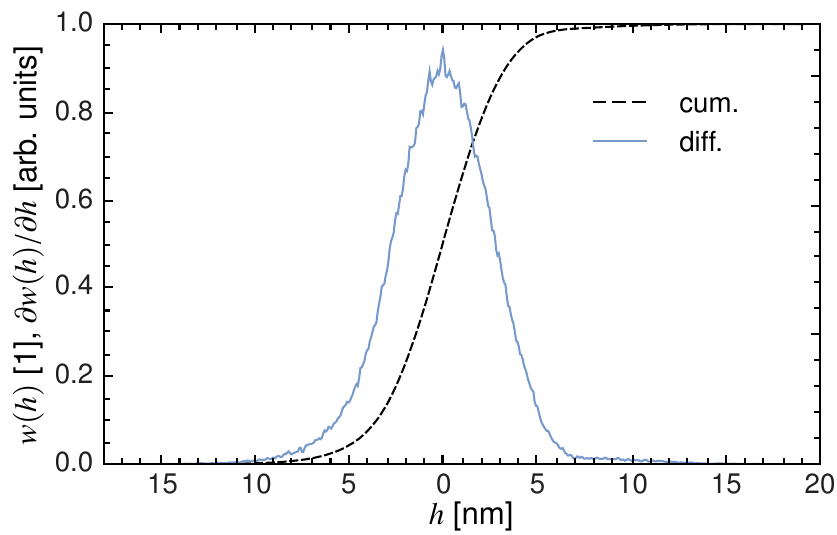}\label{fig:height_stats_hist}}
\caption{(color online) Statistical data on the scan shown in \figref{fig:4um_patches_topo}. a) correlation functions $ACF(r)$ as defined in \eqnref{eq:corr_fun}, computed from data in \figref{fig:4um_patches_topo_flat} (flattened by subtraction of $h_f$, dashed line), or after subtraction of a free polynomial of fourth order from the data in \figref{fig:4um_patches_topo_curved}(solid line), a) differential and cumulative height distribution of unfiltered data of the flattened AFM scan in \figref{fig:4um_patches_topo_flat}.\label{fig:height_stats}}
\end{figure*}
  In typical analysis found in the literature, roughness parameters are determined from AFM scans, which are flattened in order to remove the curvature (in the case of a sphere) and to compensate for piezo drift (for open loop scans). Depending on the applied method of flattening and the size $A$ of the scan area, low-frequency spectral components are eliminated from the profile $h(x,y)$ (for the sake of simplicity, we drop the index $AFM$ from now on). This latter process influences the shape of the auto-correlation function 
\begin{align}
\label{eq:corr_fun}
ACF(r\hspace{-0.7pt})\hspace{-1pt}=\hspace{-1pt}(1/\hspace{-0.7pt}A)\hspace{-1pt}\left\langle\int_A\hspace{-2pt}{\rm d}x'\,{\rm d}y'\,\|h(x\hspace{-1pt}+\hspace{-1pt}x'\hspace{-1pt},y\hspace{-1pt}+\hspace{-1pt}y')h(x,y)\|\hspace{-1pt}\right\rangle_{\hspace{-1.5pt}r}\hspace{-2pt},
\end{align}
from which $\xi$ can be obtained under the assumption of purely stochastic roughness by fitting $ACF(r)$ to the function $ACF_{\text{fit}}(r)=\G_0\exp\left[-\left(r/\xi\right)^{2H}\right]$, where $0<H<1$ is the roughness exponent, $\G_0$ is a constant, $r=\sqrt{x^2+y^2}$, and the notation $\langle\,\rangle_r$ indicates averaging over all pairs $(x,y)$ on a circle of radius $r$. 
\begin{figure}[!ht]
\centering
   \subfloat[original topology]{\includegraphics{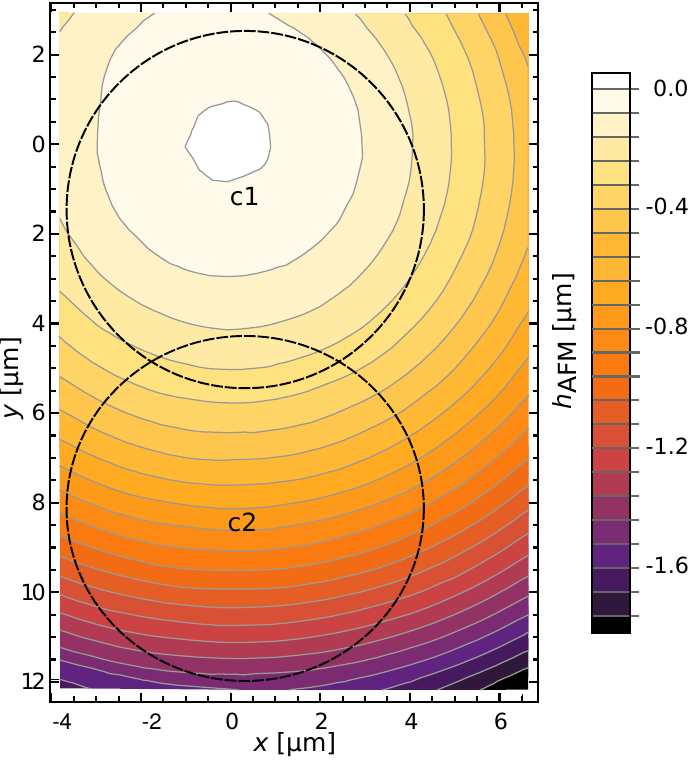}\label{fig:4um_patches_topo_curved}}\hspace{1ex}
   \subfloat[flattened topology]{\includegraphics{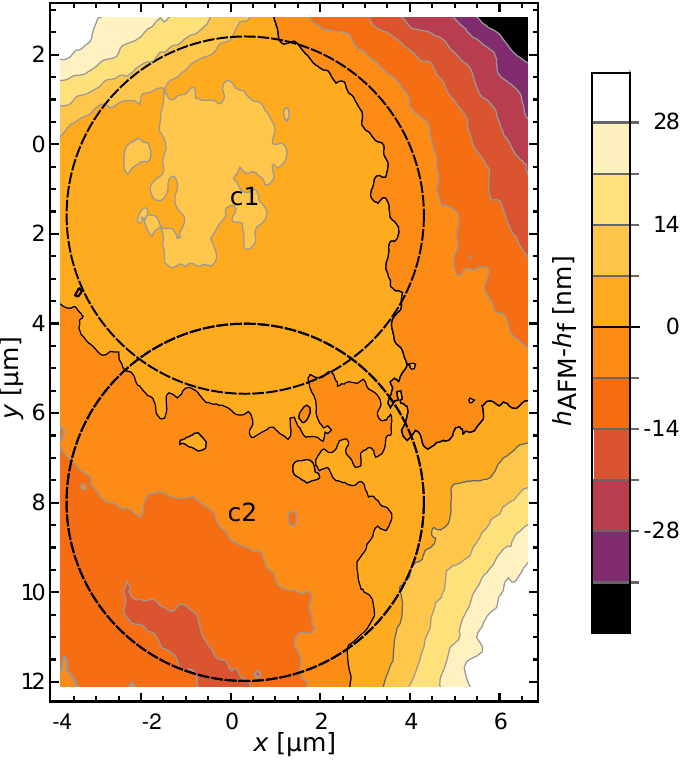}\label{fig:4um_patches_topo_flat}}
\caption{(color online) Topology of an exemplary 4310A sphere after sputtering. a) Actual AFM data ($h_{AFM}$), b) flattened by subtraction of the spherical fit $h_{f}$.\label{fig:4um_patches_topo}}
\end{figure}
\figref{fig:height_stats_ac_fun} shows an example of the function $ACF(r)$ obtained from unfiltered AFM data shown in \figref{fig:4um_patches_topo}. If the best fit of $h(x,y)$ to a sphere according to \eqnref{eq:sph_height_function} is removed from the data (resulting in the topology shown in \figref{fig:4um_patches_topo_flat}), a subsequent fit of $ACF(r)$ results in $\xi=2.5\,\m$m (dashed black line in \figref{fig:height_stats_ac_fun}). Despite the fact that the height distribution (solid line in \figref{fig:height_stats_hist}) almost perfectly follows a Gaussian curve, the spectral distribution $ACF(r)$ in \figref{fig:height_stats_ac_fun} cannot be fit satisfactorily by the model of \eqnref{eq:corr_fun}. As can be seen in the figure, a significant contribution comes from large-scale corrugations with wavelengths $\l_{corr}$ of several $\m$m~\footnote{We note that 
the large-scale deformations, which have a typical height of several nm ($4.2\,$nm(RMS), at wavelengths $>\xi=2.5\,\m$m), cannot be explained by piezo drift since all measurements have been performed in closed loop operation with accuracy better than $0.5\,$nm. Moreover we could not find any regular patterns or correlations between the irregularities shown by different scans.}, $\l_{corr}\gg\xi$.  These contributions of low spatial frequency break the conditions (1.) and (2.). Consider for example an experiment with $R=50\,\m$m, and $\xi=2.5\,\m$m. At a separation $d_0=100\,$nm roughly three structures of area $\xi^2\pi$ fit into a cap of size $r_{3\sigma,C}^2\pi$, and $R d_0/\xi^2=0.8$. Therefore, a statistical evaluation of roughness might in this case not be rectified at $d_0\lesssim1\,\m$m. It is interesting to note that, if one removes independent polynomials of fourth order in both lateral directions instead of the spherical surface $h_f$, the fit to $ACF$ leads to $\xi=420\,$nm (solid light blue line in \figref{fig:height_stats_ac_fun}). The latter value compares well with those reported in the literature ($200\,$nm in Ref.~\cite{Chen:2004} and $\lesssim 600\,$nm in Ref.~\cite{Decca:2005}).\\
\begin{figure}[!ht]
\centering
   \includegraphics{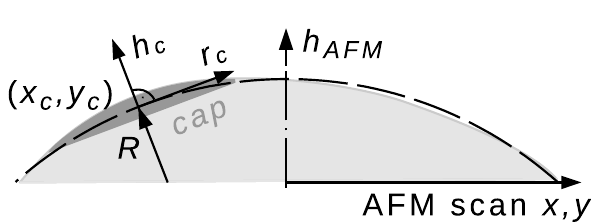}
\caption{Adaption of the local coordinate system of the cap to the tangential plane to the sphere at ($x_c,y_c$).\label{fig:patch_rot}}
\end{figure}

\begin{figure}[!ht]
\centering
   \includegraphics{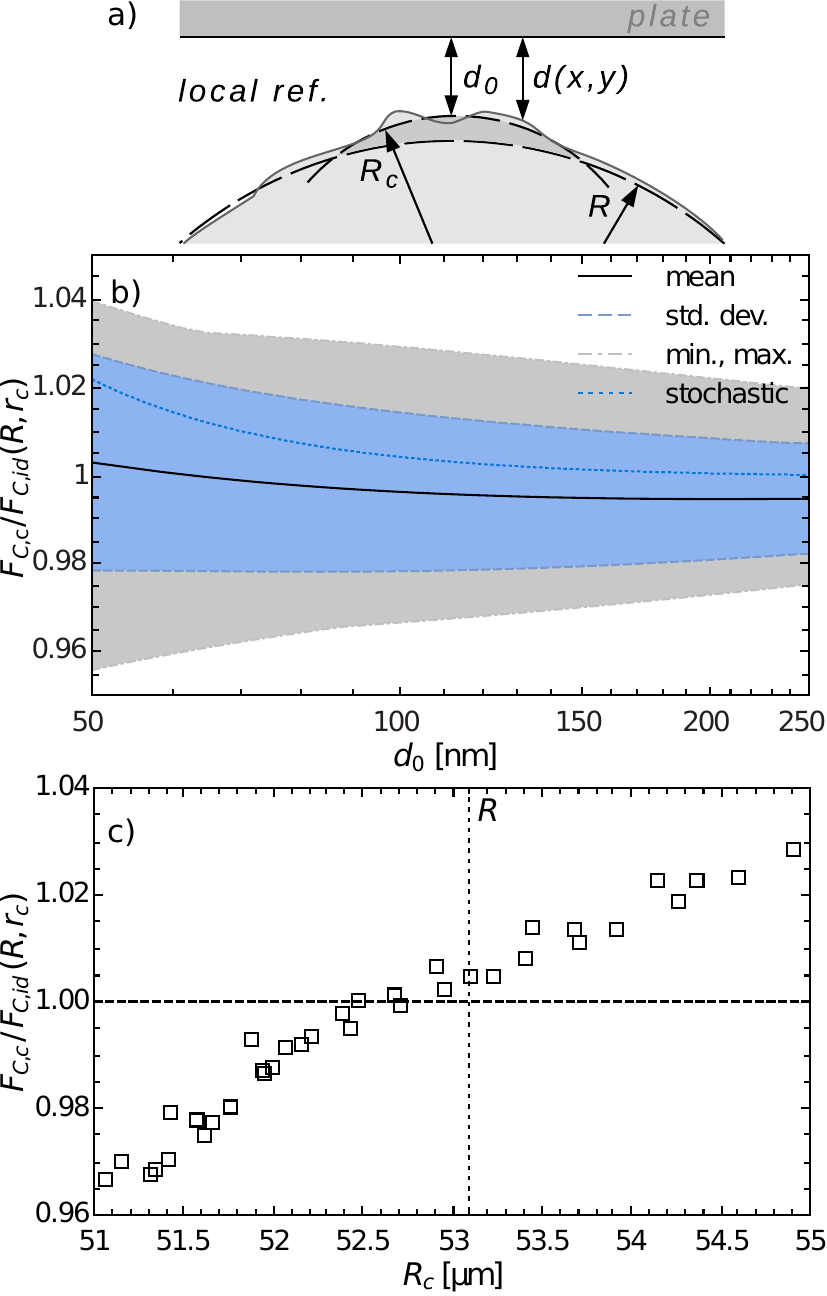}
\caption{(color online) Definitions and results for local radius reference. a) Definition of the distance reference $d_0'=d_0-h_0$ from the local spherical fit (dashed line, $R_c$), b) normalized Casimir force $F_{C,c}/F_{C,id}(R,r_c)$ as a function of $d_0'$, computed for 36 caps of $r_c=4\,\m$m on the AFM scan shown in \figref{fig:4um_patches_topo}, c) almost linear dependence of the force on the local curvature radius $R_c$.\label{fig:cap_local}}
\end{figure}

\begin{figure}[!ht]
 \centering
   \includegraphics{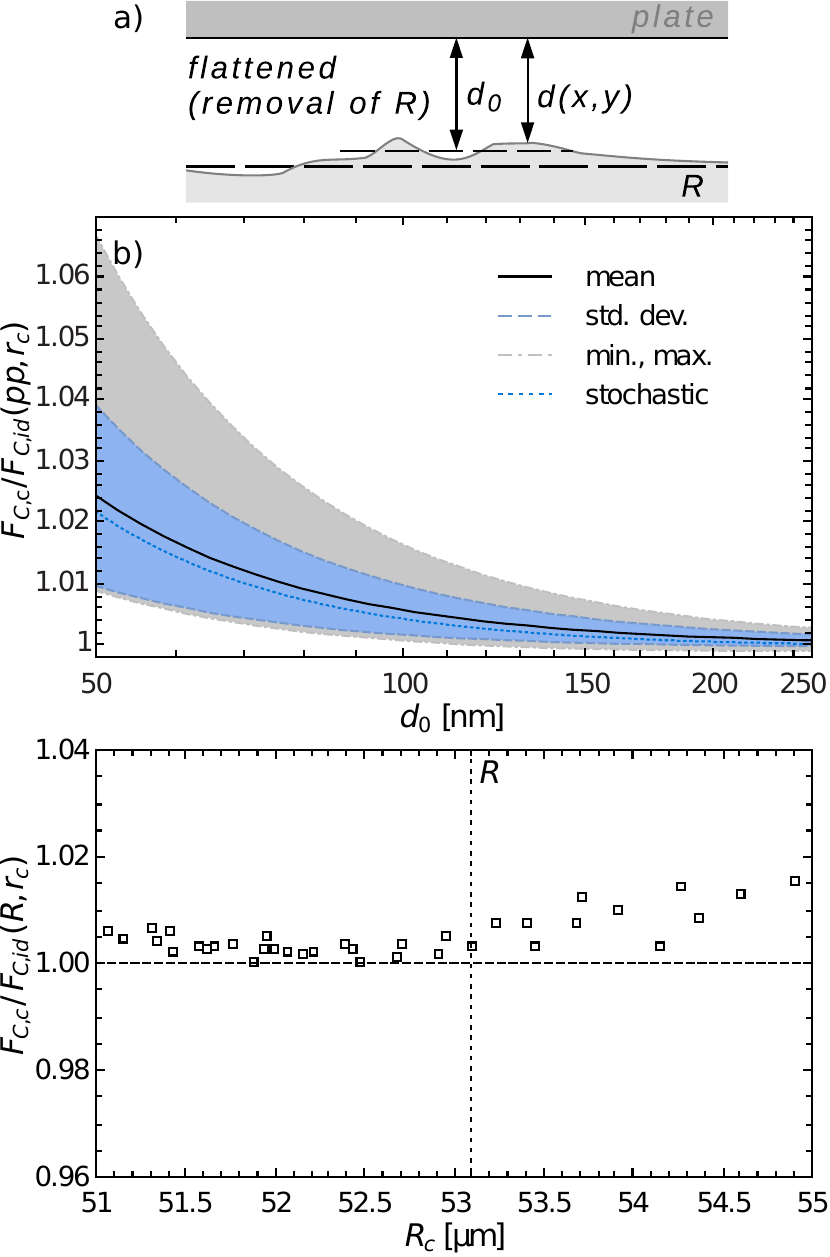}
\caption{(color online) Definitions and results for a topology flattened by removal of the global spherical fit $R$, and the local geometric mean $d_0$ as reference. a) Definition of the distance reference $d_0$ from the local geometric average of the flattened cap (short dashed line), b) normalized Casimir force $F_{C,c}/F_{C,id}(pp,r_c)$ as a function of $d_0$, computed for 36 flattened caps of $r_c=4\,\m$m on the AFM scan shown in \figref{fig:4um_patches_topo}, c) only slight dependence of the force on the local curvature radius $R_c$ after flattening of the profile.\label{fig:cap_flat}}
\end{figure}

It is further instructive to perform explicit calculations of the Casimir interaction between an ideally flat surface and the height profile $h(x,y)$ of a sphere obtained from actual AFM data. We select a representative scan, which was recorded on a 4310A sphere after sputtering. \figref{fig:4um_patches_topo_curved} shows the topology $h(x,y)$~\footnote{Note that the scan in \figref{fig:4um_patches_topo} was taken from an area of $15\times15\,\m{\rm m}^2$. The deviation of the aspect ratio from unity is a direct consequence of the scaling factor $f_D$ (here in the direction of $x$).}, where, for the purpose of a more clear illustration, a running mean filter with a lateral width of $100\,$nm has been applied. For the unfiltered data, we calculate a statistical (stochastic) roughness correction $\chi_r(d_0)$ to the Casimir force based on the established model of Ref.~\cite{Bordag:2001}, which applies the PFA. For the ease of computation we use a BSpline-interpolation of the differential height quantity $w(h)$ of the histogram in \figref{fig:height_stats_hist} (based on the flattened topography in \figref{fig:4um_patches_topo_flat}), and write
 \begin{align}
 \label{eq:pfa_roughness_corr}
  \chi_r(d_0)&=\int_{\min(h)}^{\max(h)}{\rm d}h\, w(h) \frac{F_{C,pp}(d_0-h)}{F_{C,pp}(d_0)}\,,\\
  \text{with }1&=\int_{\min(h)}^{\max(h)}{\rm d}h\, w(h)\,.\nonumber
 \end{align}
Results of \eqnref{eq:pfa_roughness_corr} will be taken as a reference for the subsequent analysis.\\
From the unfiltered data of the scan shown in \figref{fig:4um_patches_topo_curved}, we extract 36 equally distributed caps of radius $r_c=4\,\mu$m and center positions $(x_c,y_c)$~\footnote{We would like to note that the same investigations have been repeated for several different spheres. In all cases, we obtained equal qualitative (but not the same quantitative) results for all points of the presented discussion.}. For each cap $c$, the height profile $h_c$ has been transformed such that the vertical $z$-axis is parallel to the normal vector associated with the global spherical fit at $(x_c,y_c)$ (see \figref{fig:patch_rot}). The Casimir force, which would act between the curved caps with a corrugated surface profile $h_c(x,y)$ and a perfectly flat plate (in this example, both made of gold) at a distance $d_0$ can be computed by numerical application of the PFA,
   \begin{align}
 \label{eq:pfa_cap_summation}
  F_{C,c}(d_0)=\frac{A_c}{N_c}\sum\limits_n P_{C,pp}\left(d(x_{n},y_{n})\right)\,,
 \end{align}
 where $n$ runs over all $N_c$ points of the cap, $d(x,y)=d_0-h_c(x,y)$, and $A_c=r_c^2\pi$ is the area of the cap. An important aspect for the application of \eqnref{eq:pfa_cap_summation} is the definition of the reference $d_0$ for the surface separation. 
 In order to demonstrate the impact of the latter parameter, we elaborate the data in three different ways: 
\paragraph{Electrostatic distance determination.}
\begin{figure}[!ht]
\centering
   \includegraphics{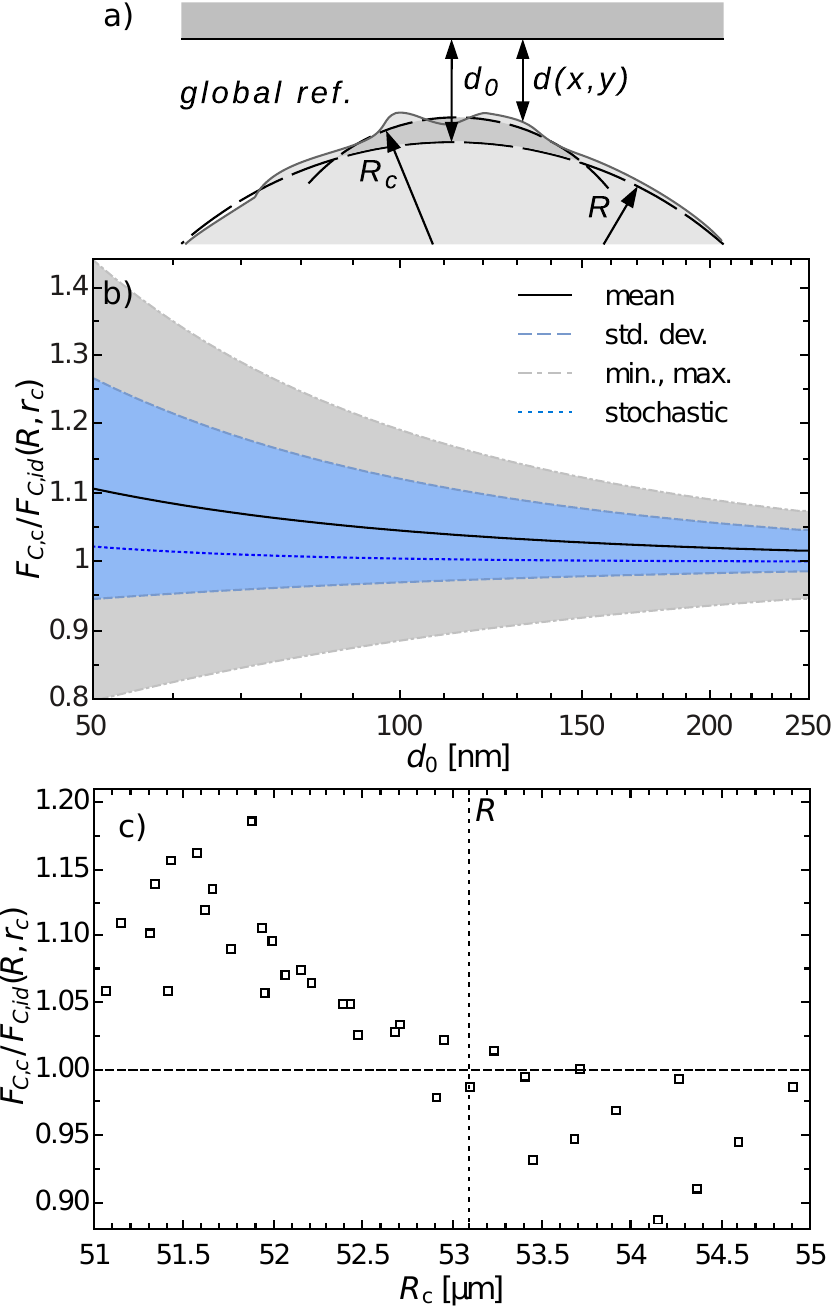}
\caption{(color online) Definitions and results for global radius reference. a) Definition of the distance reference $d_0$ from the global spherical fit (dashed line, $R$), b) results for the normalized Casimir force $F_{C,c}/F_{C,id}(R,r_c)$ as a function of $d_0$, computed for 36 caps of $r_c=4\,\m$m on the AFM scan shown in \figref{fig:4um_patches_topo_curved}, c) inverse dependence of the force on the local curvature radius $R_c$ resulting from higher tips of more convex areas. Each data point corresponds to a different cap.\label{fig:cap_global}}
\end{figure}
 In experiments where the distance between the sphere and the plate is determined via an electrostatic calibration, the surface separation $d_0$ corresponds~\footnote{The Smythe theory for the electrostatic interaction between a sphere and a plate, which is used in this analysis as well as in the literature, does not take into account local variations in the solution of the Poisson equation due to inhomogeneous conductivity (patch effects) and corrugations of the surface. We believe that this is an important point of discussion which, however, goes beyond the scope of this paper, and we restrain ourselves to the investigation of the influence of different hypothesis found in the literature regarding the surface separation.} physically to the gap between the surface of an ideal sphere, fitted to a cap of the effective radius $r_{3\sigma,E}$, and the flat plate. In the analysis of the locality of electrostatic and Casimir forces in \secref{sec:locality} (\figref{fig:r95_ddep}) it could be seen that $r_{3\sigma,E}$ is on the order of $10\,\m$m in the distance range accessible by an AFM experiment (50--$400\,$nm). This indicates that the cap size over which the electrostatic interaction is integrated (averaged), is comparable to the one of the entire scan shown in \figref{fig:4um_patches_topo}. Therefore, we define here $d_0$ between the surface of the spherical fit with curvature $R$ to the entire scan (global fit), and the ideally flat opposing plate, as depicted in \figref{fig:cap_global}a. However, the Casimir force is still (effectively) being generated within a cap of radius $r_{3\sigma,C}< r_{3\sigma,E}$ on the curved surface. When $F_{C,c}(d_0)$ is computed for each spherical cap, according to \eqnref{eq:pfa_cap_summation} one obtains the force curves shown in \figref{fig:cap_global}b.
 Since we are interested in the deviation of the forces $F_{C,c}(d_0)$ from a theoretical prediction $F_{C,id}(R,r_c,d_0)$ for a cap of the same $r_c$ on an ideally shaped sphere of radius $R$, we normalize all given results by this latter quantity. In \figref{fig:cap_global}b, the geometric mean and standard deviation over the 36 caps are indicated as well as the stochastic correction $\chi_r(d_0)$. Note that, due to the lateral separation of $1\,\m$m between the center positions of adjacent caps and the resulting overlap, the force curves are not strictly statistically independent. Nonetheless, the statistics are representative for the results in an actual experiment, in which the point of closest approach would be shifted in the same way by turning the sphere. The large spread of $\sim\pm35\,$\% in $F_{C,c}(d_0)$ at the shortest surface separation can intuitively be understood from the fact that the top surfaces of some caps lie closer to the opposing surface than others, resulting in a deviation between the effective surface separations for Casimir and electrostatic interactions. The caps indicated by `c1' and `c2' (in \figref{fig:4um_patches_topo}) correspond to the minimum and maximum force lines, respectively, in \figref{fig:cap_global}b, and represent the caps with the lowest local minimum and highest maximum values of $h_c$, respectively~\footnote{Due to the circular shape of the caps, 
the upper left and lower right corners of the scan cannot be reached, and do not enter the calculations, thus.}. An interesting fact to note is that `c1' is a convex area in \figref{fig:4um_patches_topo}, which indicates that $R_c({\rm c1})<R$, while `c2' is concave $R_c({\rm c2})>R$. The linear dependence on the curvature radius $R$, as predicted by the PFA~\cite{Blocki:1977}, is thus overruled by the influence of the distance shift, as shown in \figref{fig:cap_global}c.
\paragraph{Distance determination from contact.} If the distance is adapted to the local curvature radius $R_c$ (derived from a fit to the local spherical surface within $r_c$ around patches on the scan shown in \figref{fig:4um_patches_topo_curved}), the separation between the sphere and the plate has to be redefined as depicted in \figref{fig:cap_local}a. 
According to Ref.~\cite{vanZwol:2009} this distance reference corresponds to the one that would be obtained experimentally by moving the surfaces to contact, and subsequent consideration of the statistics of the roughness distribution measured on a flattened surface. We note that the local adaption of $d_0$ to the topology of caps of the size of $r_{3\sigma,C}$ (instead of $r_{3\sigma,E}$) naturally increases the variation in $d_0$ but reflects more accurately the situation `experienced' by the Casimir force. Accordingly, the variation in the $F_{C,c}(d_0)$ between different caps for this example (see \figref{fig:cap_local}b) drops by a factor $\sim7$ with respect to the one shown by the data in \figref{fig:cap_global}b.
As can be seen in \figref{fig:cap_local}c, the linear dependence of $F_{C,c}$ on the surface curvature $R_c$ is restored. In accordance with this finding, the minimum line in \figref{fig:cap_local}b now corresponds to `c2' and the maximum line to `c1', which is in agreement with the fact that the convex areas in \figref{fig:4um_patches_topo} have a smaller local $R_c$ (and, thus, produce a weaker force) than concave areas. Note that 
the lateral distance between `c1', and `c2' is only $~6.4\,\m$m, corresponding to an angular shift of $7\,^\circ$ on the sphere~\footnote{The same analysis on other spheres lead to shifts of less than $3\,\m$m. Also, the caps giving the minimum and maximum forces for different height reference are not necessarily the same, as in the example in the main text. The general conclusions regarding the influence of the distance error and the local curvature hold.}. Note further, that single force curves deviate not only in amplitude but also in the exponent of the distance. Therefore, the minimum and maximum lines do not necessarily correspond to actual force curves but show the min. and max. values of all curves at the same distance.\\
 \paragraph{Flattened topography, geometric average reference.}
 Finally, we investigate the influence of large-scale corrugations on the flattened topology shown in \figref{fig:4um_patches_topo}b, obtained by removal of the curvature through subtraction of the global reference $h_{f}$ (corresponding to the global radius $R$) from the AFM data. As shown in \figref{fig:cap_flat}a, the reference $d_0$ is defined by the geometric average of the local surface profile within $r_{3\sigma,C}$, according to the relation $\sum_{n=1}^{N_c} [d_0-d(x_n,y_n)]=0$\footnote{Note that the this situation of a flattened rough sphere opposing a perfectly smooth plate is equivalent to the model used to derive expressions for roughness corrections to $F_C$ for the sphere-plate geometry~\cite{Bordag:2001,Genet:2003,Broer:2011}.}. If the force\footnote{Note that $F_{C,c}$ is now normalized with respect to the force $F_{C,id}(pp,r_c)$ between ideal flat circular plates of lateral radius $r_c$.} is computed between a perfectly smooth plate at a distance $d_0$ (see \figref{fig:cap_flat}a) and caps at the same positions as before, but now on the flattened profile of \figref{fig:4um_patches_topo_flat}, the spread is again reduced slightly. More important, the change in the topology from curved to flat also leads to an up-bending of the curves at short distance in \figref{fig:cap_flat}b, which represents a qualitative change of the results with respect to the case of curved surfaces. This effect is caused by the geometric (linear) method of statistical averaging of the values $h(x,y)$. A height distribution following (according to the histogram in \figref{fig:height_stats_hist}) a Gaussian statistic around $d_0$, will naturally, due to the non-geometric distance scaling of the force, lead to a distribution of $F_C$ that is shifted towards higher values. 
 Mathematically, the average distance $d_0=\sum_{n=1}^N d(x_n,y_n)/N$ results 
in an average force $\overline{F}_C=[F_{C,pp}(d_0)/N]\sum_{n=1}^N \left(d_0/d(x_n,y_n)\right)^{3}> F_{pp}(d_0)$, where the distance dependence of the Casimir force $F_{C,pp}$ between ideal parallel plates has been approximated for the sake of simplicity by $1/d^3$. This effect of `shifting by weighting' is pronounced much stronger for the case of the flattened surfaces in \figref{fig:cap_flat} than for curved surfaces. Again, the reason is the locality of the interaction. While on caps with finite $R_c$ the central region contributes most, and only the very realization of the roughness at the point of closest approach determines the actual force, on a (flattened) cap with $R_c\to \infty$ all points contribute in the same way, and the statistical averaging enhances the force. Thus, there is a qualitative difference between the actual dependence on roughness for flat and curved surfaces.\\
Numerical evaluation of \eqnref{eq:pfa_roughness_corr} results in the blue dotted lines in Figs.~\ref{fig:cap_global}b, \ref{fig:cap_local}b, and \ref{fig:cap_flat}b. In all three figures, the statistical prediction $\chi_r$ lies within the blue (innermost) band corresponding to the standard deviation of the 36 force curves at each distance. Nonetheless, in the case of curved surfaces, $\chi_r$ shows significant deviations from the mean of the force curves obtained in our explicit example. For the computations on flat surfaces, the local averages $d_0(c)$ do not vary strongly, leading to a better agreement with $\chi_r$ (for which the average distance $d_0(\chi_r)$ is equal to the plane given by the global spherical fit indicated by $R$ in \figref{fig:cap_flat}a).  However, it has to be considered that, according to the analysis above, a purely statistical treatment may not be valid for curved surfaces. As has been indicated by other authors as well~\cite{vanZwol:2009,Klimchitskaya:1996b}, one has to consider that in an actual experiment only one of these force curves will be effective. The curves corresponding to single realizations show significant variations in both amplitude and exponent of the distance. Unless the very point of contact on the spherical surface can be determined with sub-$\m$m accuracy and the topology at this very point is taken into account in a comparison with theory, there may always be a significant deviation of analytic and experimental results.

\begin{figure*}[!ht]
\centering
   \hspace{-2ex}\subfloat[4310A]{\includegraphics{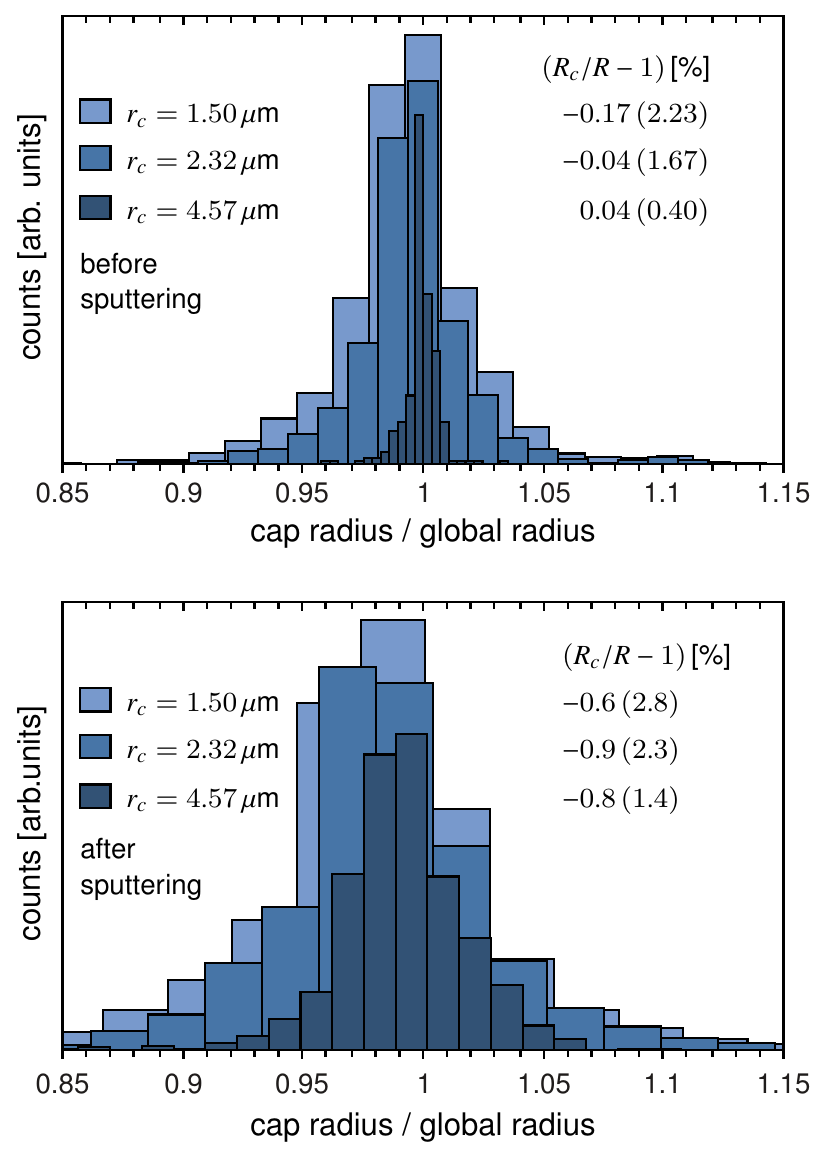}\label{fig:stat_distr_100um}}\hspace{0.1ex}
   \subfloat[4320A]{\includegraphics{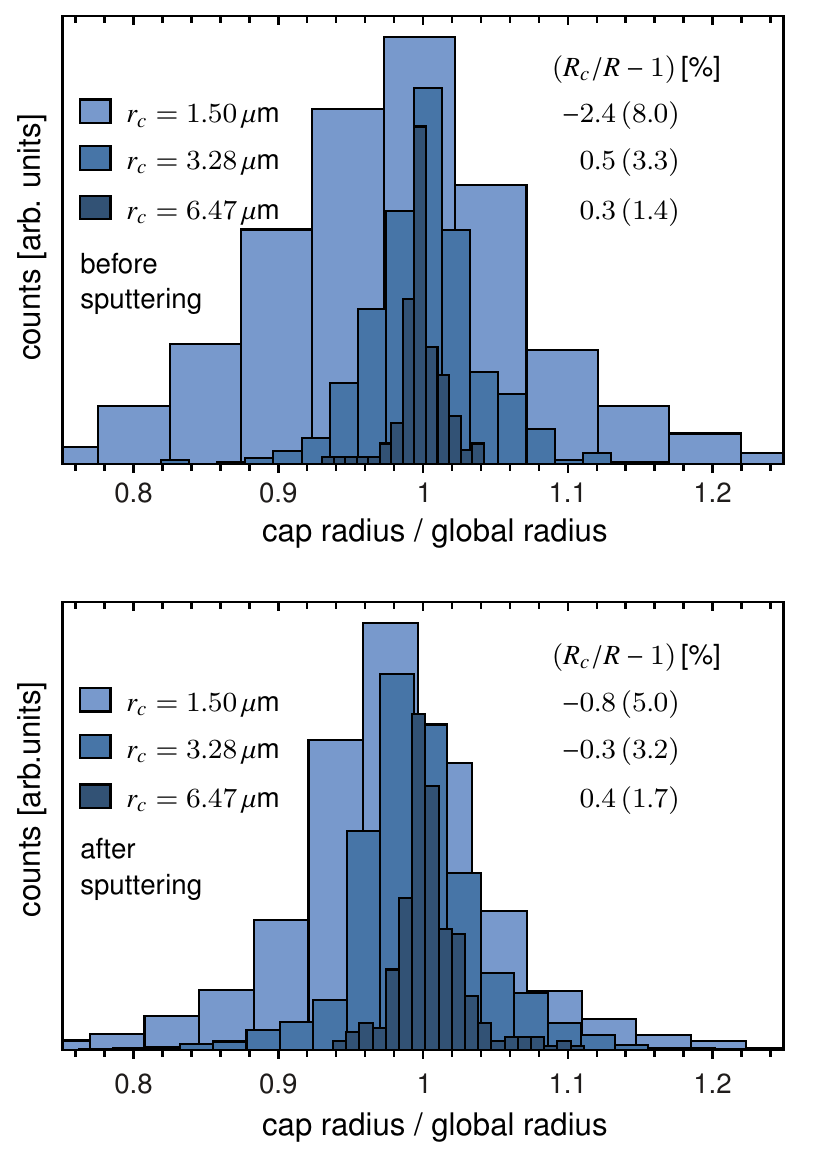}\label{fig:stat_distr_200um}}
   \caption{(color online) Histograms of the fitted local radii on spheres of type 4310A (a) and 4320A (b), before (upper figures) and after (lower figures) sputtering the surfaces. In order to be comparable, the cap radii have been normalized by the globally fitted radius for each sphere (1 on the ordinate). The width of the histogram bars has been set to $\sigma_R/2\,$, where $\sigma_R$ is the standard deviation of the respective (assumed Gaussian) distribution.\label{fig:stat_distr}}
\end{figure*}

The previous qualitative investigation was based on a single (representative) AFM scan and can therefore not be used to derive a quantitative estimate of the error due to local variation of the surface topology. In order to achieve the latter, we assume that the distance is determined as outlined in Ref.~\cite{vanZwol:2009}, which corresponds to the case that the height reference is obtained from the local spherical fit. Then, the variation of the expected Casimir force scales directly with the curvature radius (see~\figref{fig:cap_local}c). We have performed fits using \eqnref{eq:sph_height_function} to caps of different $r_c$ for all available unfiltered AFM data. 
The results are shown as histograms in \figref{fig:stat_distr}. For each sphere, the $R_c$ have been normalized with respect to the corresponding global fit radius $R$. Numerical values for the weighted mean $\overline{R}_c$ and standard deviation $\sigma_c$ of each distribution are given in tabulary form in the figures. Throughout, the agreement between $\overline{R}_c$ and $R$ is very good, which rules out systematic effects.\\
Considering variations in the Casimir force, the most important parameter of this analysis is $\sigma_R(r_{3\sigma,C})$ which, averaged over all spheres of type 4310A for $r_c=r_{3\sigma,C}$ at $d_0=100\,$nm, gives a value $1.4\,$\% as compared to the standard deviation $1.8\,$\% of the 36 caps with $r_c=4\,\m$m used in the evaluation of the scan shown in \figref{fig:4um_patches_topo}. On 4320A colloids, $\sigma_R(r_{3\sigma,C})$ takes a similar value of $1.7\,$\%.
The widths $\sigma_R$ of the (fitted Gaussian) distributions, naturally increase for smaller cap sizes (corresponding to smaller surface separations) and approach, in the limit $r_c\to 0$ the respective value of the roughness evaluation using the full resolution of the AFM scans.
However, one has to bear in mind that these statistical measures do not apply in the comparison of force measurements taken between a single spot on a sphere, and a smooth plate. As discussed above, an actual experiment will be influenced by the local realization of the geometry of the surface, and the resulting uncertainty may be much larger than $\sigma_R$. We would like to point out, that this potential error is neither covered by a typical statistical analysis based on a stochastic distribution of roughness nor by consideration of the uncertainty in $R$ determined from SEM images. The resolution of the latter is clearly insufficient to quantify surface irregularities with a height of the order $10\,$nm.\\
Finally, we note that the present investigation is focused on the experimentally relevant situation of a micro-sphere opposing a flat plate, where at least one of the interacting surfaces has a finite curvature. In the case of parallel plates (as for example in mirco-electromechanical devices), the area of interaction is much larger. Therefore, the conditions (1) and (2) mentioned at the begin of \secref{sec:results_locality} are met, and the effect of surface corrugations can be modeled satisfactorily statistically.
%
%
\section{Conclusion}
\label{sec:concl}
We have investigated the radius and surface roughness for colloid particles of type 4310A and 4320A, which are widely used in experiments on hydrodynamic and Casimir interactions in the sphere versus plate geometry. Numerical analysis shows that on curved surfaces only a small fraction of the surface area effectively contributes to the measured Casimir and electrostatic forces. On the basis of actual AFM topology data, our study indicates that
\begin{itemize}
 \item surfaces of colloid particles show non-regular corrugations on intermediate length scales (1--10$\,\mu$m) of comparable or larger amplitude than short scale roughness. Flattening procedures, which are necessary to measure roughness distributions, may filter these irregularities.
 \item in the presence of such corrugations, the translational invariance and the statistical representativity of the interacting `cap' area may be violated at surface separations up to a few $\m$m. Therefore, the fundamental assumptions for a statistical analysis may not hold.
 \item forces, which are measured on different spots of the same surface vary in both amplitude and exponent of the distance dependence. The scale of these variations depends on the applied reference for the distance.
 \item the effective length scale for electrostatic interactions is roughly 3 times larger than the one for the Casimir force. Thus, the quantum-mechanical force `sees' a much smaller area on the curved surface than does the electrostatic one. In the presence of corrugations at intermediate length scales, this may lead to different offsets and definitions for the distance of the two interactions, which may result in a spread in measured Casimir forces of several tens of percent.
 \item in the case that the distance is determined from the contact of the two surfaces, the error which would appear due to an electrostatic calibration (due to different effective offsets for the surface profile for electrostatic and Casimir interactions) may be avoided. This is particularly true if the plate is smooth enough (atomically flat) on \emph{all} length scales so that its roughness profile does not influence the distance at contact. Then, the local variation in the Casimir force is mainly determined by deviations of the surface curvature radius from its global value, which amounts to 1--5$\,$\%, depending on the distance. These geometrical variations, again, are to be seen on the `locality scale' of the interaction and can only be determined from unfiltered AFM topography data taken at the very point of the force measurement.
\end{itemize}
The present work shows that commonly used micro-spheres may feature surface corrugations which influence short-scale surface interactions locally. The very realization of the geometry at the point of closest approach determines the force quantitatively and qualitatively. A theory which is compared to data from such experiments must take into account the real topography of the surface, as statistical measures may not be sufficiently accurate.
\section*{Acknowledgments}
  This work was partially funded by the Foundation for Fundamental Research on Matter (FOM), which is financially supported by the Netherlands Organisation for Scientific Research (NWO). R. Sedmik acknowledges his FWF Schr\"odinger fellowship J3050-N20.

\end{document}